\documentclass[letterpaper]{article}

\usepackage{aaai}
\usepackage{comment}
\usepackage{courier}
\usepackage{graphicx}
\usepackage{helvet}
\usepackage{times}
\usepackage{url}
\usepackage{pgfgantt}
\usepackage{lipsum} 
\usepackage{caption}%
\usepackage{subcaption}
\usepackage{tabularx,booktabs}
 \usepackage{capt-of}
\newcolumntype{C}{>{\centering\arraybackslash\hsize=.5\hsize}X} 
\newcommand\citeName[1]{\citeauthor{#1}~\shortcite{#1}}

\makeatletter
\newcommand*{\centerfloat}{%
  \parindent \z@
  \leftskip \z@ \@plus 1fil \@minus \textwidth
  \rightskip\leftskip
  \parfillskip \z@skip}
\makeatother

\usepackage{amsmath}
\usepackage{algorithm}
\usepackage[noend]{algpseudocode}
\makeatletter
\def\BState{\State\hskip-\ALG@thistlm}
\makeatother

\algnewcommand\algorithmicforeach{\textbf{for each}}
\algdef{S}[FOR]{ForEach}[1]{\algorithmicforeach\ #1\ \algorithmicdo}

\usepackage[textsize=tiny]{todonotes}
\begin{document}
%
\title{Exploring Stereotypes and Biased Data with the Crowd}

\author{Zeyuan Hu \\
Department of Computer Science \\
The University of Texas at Austin \\
iamzeyuanhu@utexas.edu
\And Julia Strout \\
Department of Computer Science \\
The University of Texas at Austin \\
jstrout@utexas.edu
}

\maketitle


\section{Introduction} 
\label{section:introduction}
In 2016, Baidu and Google spent somewhere between twenty and thirty billion dollars developing and acquiring artificial intelligence and machine learning technologies \cite{McKinsey2017}. A range of other sectors, including health care, education, and manufacturing, are also predicted to adopt these technologies at increasing rates. Machine learning and AI are proven to have the capacity to greatly improve lives and spur innovation. However, as society becomes increasingly dependent on these technologies, it is crucial that we acknowledge some of the dangers, including the capacity for these algorithms to absorb and amplify harmful cultural biases. 

Algorithms are often praised for their objectivity, but machine learning algorithms have increasingly made news for a number of problematic outcomes, ranging from Google Photos incorrectly classifying African Americans as gorillas to the judicial system using algorithms that are biased against African Americans \cite{NYTimes,ProPublica}.  These harmful outcomes can be traced back to the data that was used to train the models.

Machine learning applications put a  heavy premium on data quantity. Research communities generally believe that the more training data there is, the better the learning outcome of the models will be \cite{halevy2009unreasonable}. This has led to large scale data collection. However, unless extra care is taken by the researchers, these large data sets will often contain bias that can profoundly change the learning outcome. Even minimal bias within a data set can end up being amplified by machine learning models, leading to skewed results. Researchers have found that widely used image data sets imSitu and MS-COCO, along with textual data sets mined from Google News, contain significant gender bias \cite{zhao2017men,bolukbasi2016man}. This research also found that training models with this data amplified the bias in the final outcomes. 

Once these algorithms have been improperly trained they can then be implemented into feedback loops where systems ``define their own reality and use it to justify their results'' as Cathy O'Neil describes in her book \textit{Weapons of Math Destruction}. O'Neil discusses problematic systems like PredPol, a program that predicts where crimes are most likely to occur based on past crime reports, which may unfairly target poor communities.

It therefore becomes necessary to consider the bias that may be introduced as a data set is being collected and to attempt to prevent that bias from being absorbed by an algorithm. We propose using the crowd to help uncover what bias may reside in a specific data set.

The crowd has potential to be useful for this task. One of the key difficulties in preventing bias is knowing what to look for. The varied demographics of crowd workers provide an extended range of perspectives that can help uncover stereotypes that may go unnoticed by a small group of researchers. Some work has already been conducted in this area, and \citeauthor{bolukbasi2016man} \shortcite{bolukbasi2016man} found that the crowd was useful in determining the level of stereotype associated with an occupation, as well as determining incorrectly gender biased words by asking the crowd to rate analogies such as ``she is to sewing as he is to carpentry''. We want to extend our analysis to stereotypes beyond gender, including those surrounding race and class.

The goal of our research is to contribute information about how useful the crowd is at anticipating stereotypes that may be biasing a data set without a researcher's knowledge. The results of the crowd's prediction can potentially be used during data collection to help prevent the suspected stereotypes from introducing bias to the dataset. We conduct our research by asking the crowd on Amazon's Mechanical Turk (AMT) to complete two similar Human Intelligence Tasks (HITs) by suggesting stereotypes relating to their personal experience. Our analysis of these responses focuses on determining the level of diversity in the workers' suggestions and their demographics. Through this process we begin a discussion on how useful the crowd can be in tackling this difficult problem within machine learning data collection. 

\section{Related Work}
\label{section:related-work}

\subsection{Work on bias in data sets and amplification}
As biased data sets get more coverage in the news, an increasing amount of research has been conducted around determining if data sets are biased and trying to mitigate the found bias. 

\citeauthor{torralba2011} \shortcite{torralba2011} presented a survey of data sets in the computer vision field to address the potential bias within them. Their strategy was to see how well a model trained on one data set is able to generalize and be used with alternative but similar data sets. The idea behind this is straightforward. If a data set is a truthful representation of the visual world, then a model trained on it should be able to work with other data sets in the same domain. The authors use six well known object recognition data sets and find that none are able to produce models that generalize well beyond their own data set. In this work, the researchers attribute the problems to bias in how images are selected, photographed, and problems with standardized labeling. They do not discuss the issues of cultural bias that we are focusing on in this paper, however they do bring up critical issues in how data sets are curated.

\citeauthor{zhao2017men} \shortcite{zhao2017men} explored the imSitu and MS-COCO data sets and determined that each exhibited significant gender bias. ImSitu is a situation recognition data set containing over 125,000 images labeled based on the action occurring \cite{yatskar2016situation}. The images have sub-labels that include the actors, objects, substances, and locations of the image and the roles they play. When the authors explored the imSitu data set they found that ``46.95\% of verbs favor a gender with a bias of at least 0.7,'' with words like ``shopping'', ``microwaving'', and ``cooking'' strongly biased towards the female gender and words like ``driving'', ``shooting'', and ``coaching'' biased towards the male gender. When a model was trained on this data set, 47.5\% of the verbs saw a mean amplification of 0.05\%, with originally heavily biased verbs seeing even more amplification \cite{zhao2017men}. 

\citeauthor{zhao2017men} \shortcite{zhao2017men} found similar bias in MS-COCO, a large multilabel object classification data set developed by Microsoft \cite{LinMBHPRDZ14}. One third of the noun objects were heavily biased towards men, with sports related terms showing strong bias towards men while kitchen objects were strongly biased towards women. A model trained on this data set also amplified the bias, with some of the originally more biased terms increasing bias by as much as 0.1\%. They developed a framework to reduce the bias by constraining the bias in a model to match the bias in the training data. While this strategy makes great strides in preventing bias amplification, it does not address correcting the original bias existing in the data.

\citeauthor{bolukbasi2016man} \shortcite{bolukbasi2016man} investigated the gender bias and stereotypes within word embeddings gathered from Google News, and also found significant gender bias. Their work addresses the difficult problem of maintaining gender bias for words like “queen” and “king” while removing bias from words that should be gender neutral, like “nurse” and “engineer,”  but have gender stereotypes attached to them. They provide a methodology to modify the word embeddings to remove cultural stereotypes.  

Complementary to the work of \citeauthor{bolukbasi2016man} \shortcite{bolukbasi2016man} is that of \citeauthor{Caliskan183} \shortcite{Caliskan183} who demonstrate that word embeddings generated from 2.2 million distinct words mined from crawling the Internet encode human like biases, and recommend caution to other researchers as it becomes clear that technologies can perpetuate harmful cultural stereotypes. In this work the authors reference and reproduce some of the results from the 1998 work of \citeauthor{greenwald1998measuring} \shortcite{greenwald1998measuring}, which introduced the implicit association test (IAT) and other important IAT findings since then. Some of the results they reproduce are that traditionally white names have more positive associations than traditionally African American names, young people’s names have more positive associations than old people’s names, and that math is more associated with men while arts are more associated with women. 

\citeauthor{feldman2015certifying} \shortcite{feldman2015certifying} use the legal term disparate impact, ``which occurs when a selection process has widely different outcomes for different groups, even as it appears to be neutral" to define unintentional bias. They present a variety of methods to account for disparate impact in machine learning models, including a strategy to repair the data. However, they find that sacrifices must be made in the amount of data repaired in order to keep classification rates accurate. We, along with many other researchers, share this concern about sacrificing accuracy levels. We live in a biased world, and sometimes in order to model this world, using data that incorporates cultural stereotypes can be beneficial to classification rates. This introduces a choice for individual researchers to make based on their given task. Should researchers model the world as it is or strive for a more egalitarian ideal? We do not have the answers to how bias should be handled in every case. However, as in the instances we mentioned earlier, biased algorithms are having harmful impact on individual’s lives and it is necessary for the research community to try an address them. 

\subsection{Demographics of the crowd}

The biggest motivation for us to use crowdsourcing to uncover the bias in the dataset is the diversity of the sample
drawn from the crowd \cite{Berinsky2012}.Thus, it is helpful to 
know the demographics of the crowd first before we can leverage the power of the crowd to uncover any hidden bias residing in the dataset. We use AMT as our crowdsourcing platform. Since Amazon
does not disclose any identity information (i.e., language abilities, residence country), which is key to understanding
the demographics of the crowd, researchers have carried out various studies to understand the crowd better. \citeName{Pavlick14} performed a large scale study of the language abilities of the
workers on AMT. Specifically, they target bilingual workers and give a list of self-reported native languages of the 
workers. In addition, they find that there are 13 languages (Dutch, French, German, Gujarati, Italian, Kan- nada, Malayalam, 
Portuguese, Romanian, Serbian, Spanish, Tagalog, and Telugu) that have large worker populations and workers who speak these 
languages can complete tasks (i.e., translations) quickly with good quality. 
Getting a sense of the languages spoken by the workers on AMT is crucial for our research
because studies \cite{panos} have shown that people who speak different languages tend to view the world differently and
bias is naturally inherited in the languages. 

\citeName{Ipeirotis2010} studied the workers on AMT from their demographics (i.e., age, gender, martial status) 
and social backgrounds (i.e., income levels, motivation as a worker).
He finds out that about 75\% of the workers they surveyed are the U.S. residents. 40\% of them are younger
than 30 and about 60\% of the respondents are female. In addition, 70\% hold Bachelor degrees or higher and 60\% have
annual income greater than \$25,000 per year. \citeName{Ross2010} extends \citeauthor{Ipeirotis2010}'s work 
and find that the majority of the workers on AMT has shifted from the U.S. to the India, which counted towards 39\% and
46\% of the worker population respectively. 

The diverse demographic information of the crowd is also confirmed by various studies, in which they use the crowdsourcing
as a way to access the rare or hard-to-reach populations. \citeName{Duncan2003} studied illicit drug users using
Internet-based surveys and \citeName{SSQU12} carried out a study involving LGBT group using Internet-based approach.
Those studies make us believe that an online approach via AMT can help us understand bias from people with different cultural
and social backgrounds.

\subsection{Prior Crowd work}
We are not the first to consider using the crowd to explore the problem of bias in data. As mentioned above, \citeauthor{bolukbasi2016man} found significant gender bias in word embeddings gathered from Google News. They also employed the crowd in their research. They sent a total of three questionnaires to the crowd. First, they asked the crowd to think of sets of words that were either related to a gender by definition or by stereotype. Next, they asked the workers to fill in an analogy reflecting gender stereotypes, and the final questionnaire asked them to evaluate an analogy for the amount of stereotyping it exhibited. They found that ``the geometric biases of embedding vectors is aligned with crowd judgment of gender stereotypes'' \cite{bolukbasi2016man}. 

When soliciting stereotypes from the crowd, \citeauthor{bolukbasi2016man} \shortcite{bolukbasi2016man} narrowed their scope to only asking about occupations as, ``they are easily interpretable by humans and often capture common gender stereotypes.'' We will expand our focus to stereotypes outside of gender, as well as stereotypes beyond occupation.

\citeauthor{attenberg2011beat} \shortcite{attenberg2011beat} conducted work in using the crowd to find the ``unknown unknowns" in trained model. While their work does not relate specifically to bias, we are also searching out ``unknown unknowns" in our data set, in this case the stereotypes or bias that are reflected in a data without our knowing it. In their research the authors found that by creating a game to incentive workers to think up test cases that would stump the model, they were able to uncover problems that they did not realize existed in their model. We will similarly be asking the crowd to help us think of what may be missing in a data set. 

\citeauthor{durupinar2016environment} \shortcite{durupinar2016environment} asked the crowd about stereotypes surrounding personality, nationality, and profession in order to more accurately animate humans in games. Three hundred and sixty three AMT workers were asked to scale how likely a person of a given nationality or profession was to exhibit certain personality traits. The results demonstrated that workers ``do have stereotypes about the expected personality of people from given nationalities and professions''\cite{durupinar2016environment}.

\section{Methodology}
\label{section:method} 

\subsection{Questionnaire 1}
Our task design is focused around asking the crowd to suggest stereotypes that may be introducing bias into a data set. We frame our experiment around the general structure of a situation recognition data set that would include images of various types of people participating in a range of activities. In the first iteration of our questionnaire sent to the crowd, we asked for suggestions about types of stereotypical situations within the categories of gender, race, and class. 

\begin{figure*}[p!]
\centering
\label{Q1}
\includegraphics[scale=0.5]{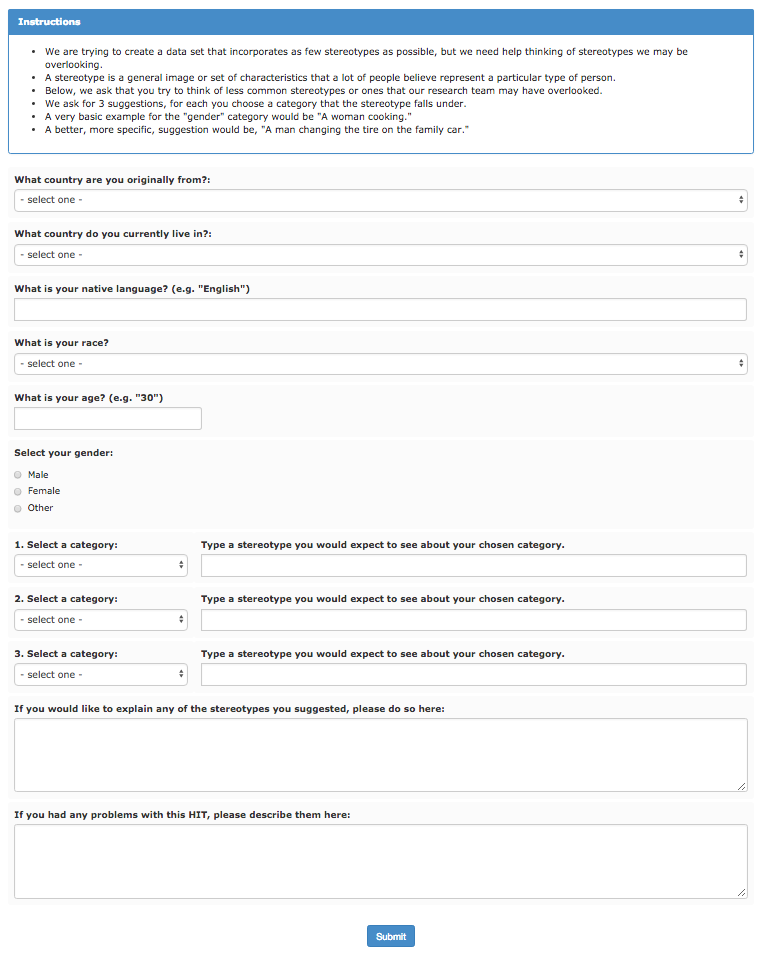} 
\caption{Questionnaire 1}
\end{figure*}

The first questionnaire sent to the crowd is shown in Figure 1. The format of this questionnaire was to first ask the worker a few demographics questions and then have them suggest three different stereotypes. The demographic questions are:
\begin{itemize}
\item What country are you originally from?
\item What country do you currently live in?
\item What is your native language?
\item What is your race?
\item What is your age?
\item What is your gender?
\end{itemize}

After these demographics we ask them to think of stereotypes that a research team may have overlooked. For each stereotype they are instructed to choose a category, with the options being ``Race", ``Gender", ``Class", and ``Other," and then to type a stereotype in a free form text box. The HIT also includes two optional text boxes, one where workers can explain anything about their suggestions, and another where they can comment on any problems with the HIT.  

\begin{figure*}[p!]
\centering
\label{Q2}
\includegraphics[scale=0.5]{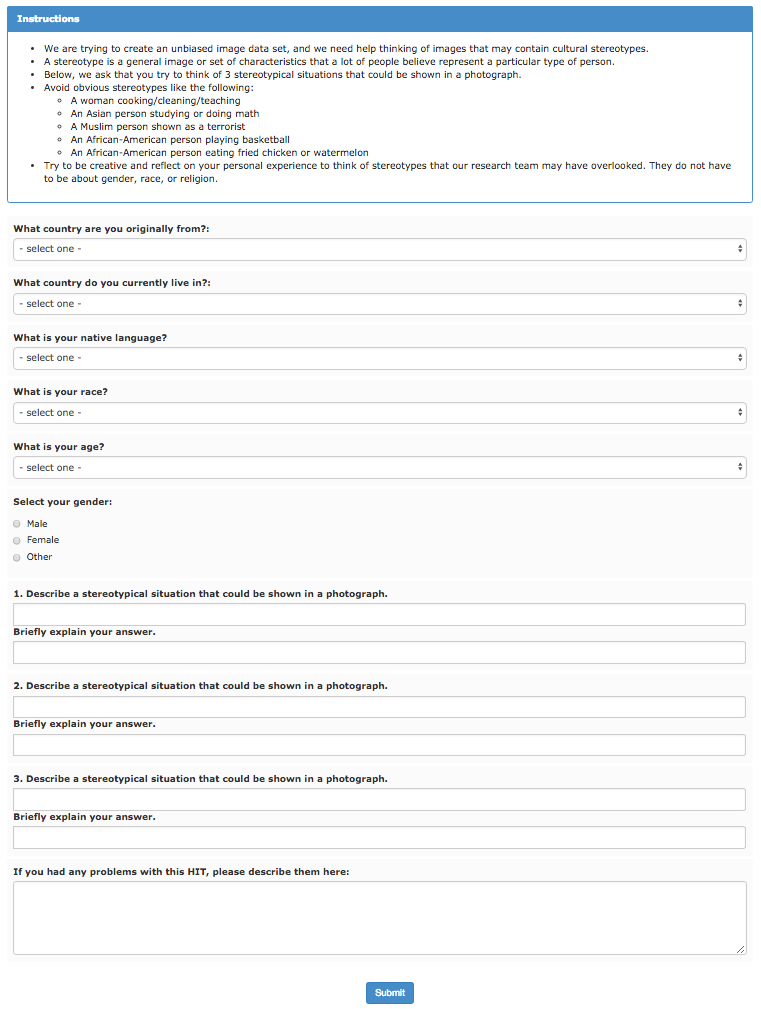} 
\caption{Questionnaire 2}
\end{figure*}

\subsection{Questionnaire 2} When we initially planned to conduct our research in two stages, the plan was that we would the analysis of the first questionnaire to limit the scope of the second questionnaire by restricting the type of stereotypes we asked about or using selection rather than free response questions. Instead, the most important update we had to make in round two was to the instructions which, when reevaluated, had been far too ambiguous in the first HIT. They resulted in vague stereotypes from the crowd that were often only one word, and rarely a scenario like we had intended. In \emph{Questionnaire 2}, the instructions were updated to specifically state that the suggested stereotypes should be situations that could be represented in a photograph. 

Instead of limiting the scope of the crowd suggestions, we also decided to widen it by removing the category labels. Although we had included an ``Other" category with the hopes that workers would think of creative suggestions outside of race, class, or gender, the overwhelming majority of suggestions from \emph{Questionnaire 1} were in those three predefined categories. We removed the category selection and instead asked each worker to briefly describe their suggestion with the hopes it would cause them to produce better quality results.

Analyzing the results of the first task also prompted changing the free text response format of the some of the demographic questions to be all drop down menus to remove the need for data cleaning. In the first task, the US and India had been the first two available countries in the drop down country list. To ensure that workers weren't just choosing the first items in the list, they were returned to alphabetical order.

\subsection{HIT Logistics}
Each HIT paid a worker \$0.05. We requested 200 HITs for each questionnaire, and received 174 responses within three days from the first and 148 within four days from the second. We gave workers 30 minutes to complete the task.

\section{Analysis}
\label{section:analysis1}

\subsection{Questionnaire 1 and 2 Dataset Quality}
\begin{figure}[!b]
\centering
\includegraphics[scale=0.4]{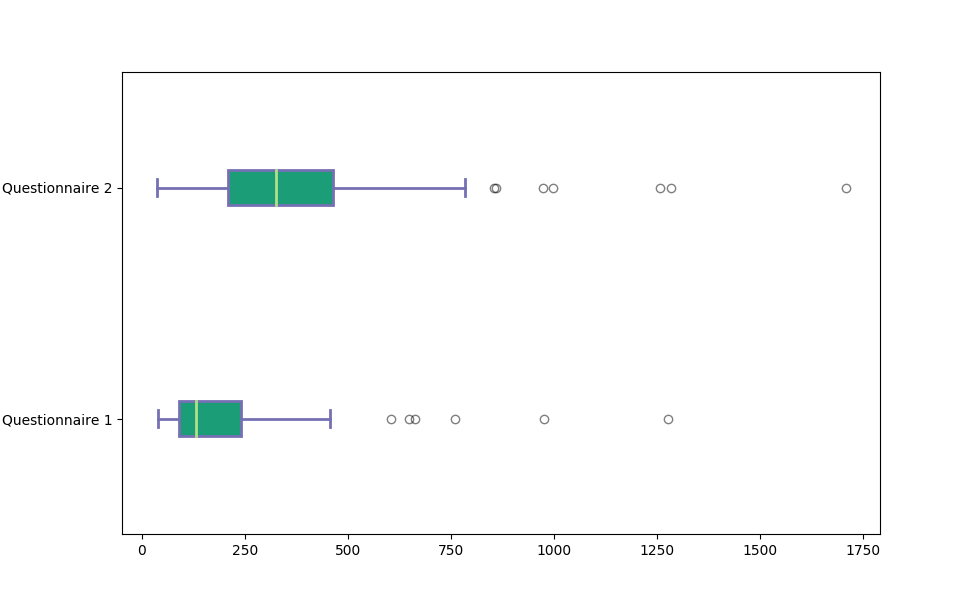}
\caption{Box plot of Task Time}
\label{fig:boxplot}
\end{figure}

174 workers answer the \emph{Questionnaire 1} and 148 workers answer the \emph{Questionnaire 2}. 
Before performing any analysis, we need to first measure the dataset quality. 
Table \ref{table1} shows the basic statistics about the dataset.
One important measure of the work quality is the task time by workers \cite{Rzeszotarski:2011}. 
Figure \ref{fig:boxplot} shows the boxplot of the task time. The average task time for \emph{Questionnaire 1} is 187.70 seconds. 
The shortest task time lasts only 39 seconds and the longest work can take 1278 seconds. The standard deviation is 161 seconds.
However, for \emph{Questionnaire 2}, the average task time is 376.89s. Workers on average spend
more time on \emph{Questionnaire 2} than \emph{Questionnaire 1}. Another important
metric to measure the dataset quality is the length of the response. If a worker spends a decent
amount of time on the task, we expect the length of his response will not be short. We use three words to define 
the shortness. In other words, if a response length is shorter than three words, 
we consider it as a suspicious response indicating that the workers may not put enough effort. 
The heuristic for picking three words is that if a high quality response should be structured as a complete sentence, which should
at least have \emph{subject}, \emph{verb}, and \emph{object}.
Under this standard, among $522$ responses, 16\% (85) of them are suspicious answers. Table \ref{table2} shows some of
the suspicious answers from \emph{Questionnaire 1} under our checking method. 
Among 85 suspicious answers, 4 of them coming from \emph{Other}, 32 of them coming from \emph{Gender}, 
28 of them coming \emph{Race}, and 21 of them coming from \emph{Class}. In addition to 
the sentence length check, further examination shows that copy-paste issue exists for lazy workers. For example,
a worker copies ``NO IDEA" three times across all three responses. We apply the same method to \emph{Questionnaire 2}
and we find that among 888 responses, only 5.4\% of them are suspicious answers, which suggests that 
workers put more effort on \emph{Questionnaire 2} than \emph{Questionnaire 1}. 

\begin{table}
\captionsetup{size=footnotesize}
\caption{Statistics from \emph{Questionnaire 1}} \label{table1}
\footnotesize\centering
\smallskip 
\begin{tabular*}{\columnwidth}{@{\extracolsep{\fill}}lcc}
\toprule
  Description  & \emph{Questionnaire 1} & \emph{Questionnaire 2} \\
\midrule
Number of completed tasks & 174 & 148     \\
HIT completion rate & 87\% &74\%\\
Average task time (seconds) & 187.70 & 376.89 \\
Average worker age & 34.64 & 35.54\\
Number of responses in \emph{Other} & 25  & 120\\
Number of responses in \emph{Gender} & 212 & 105\\
Number of responses in \emph{Race} & 170 & 157\\
Number of responses in \emph{Class} & 115 & 25\\
\bottomrule
\end{tabular*}
\end{table}

\begin{table}
\centering
\caption{Some of suspicious answers from Questionnaire 1}\label{table2}
\begin{tabular}{ |p{1.8cm}|p{2cm}|p{1.5cm}|p{2.3cm}|   }
 \hline
 \multicolumn{4}{|c|}{\textbf{Suspicious answers from \emph{Questionnaire 1}} } \\
 \hline
 \textbf{Class} & \textbf{Gender} & \textbf{Race} & \textbf{Other}\\
 \hline
 1st class   & female & none & Dumb blonde  \\ 
 dumb & gay people & lazy & Cheap Jew \\
 High class& emotional & Asian & College degree \\
 middle & NO IDEA& Hispanic & dresss \\
 \hline
\end{tabular}
\centering
\end{table}

\begin{table}
\centering
\caption{Some of suspicious answers from Questionnaire 2}
\begin{tabular}{ |p{1.8cm}|p{2cm}|p{1.5cm}|p{1.7cm}|   }
 \hline
 \multicolumn{4}{|c|}{\textbf{Suspicious answers from \emph{Questionnaire 2}} } \\
 \hline
 \multicolumn{4}{|c|}{httpswwwsurveymonkeycomr5PSZCCY} \\
 \multicolumn{4}{|c|}{experiment} \\
 \multicolumn{4}{|c|}{NICE} \\
 \multicolumn{4}{|c|}{experiment} \\
 \multicolumn{4}{|c|}{same situation}\\
 \multicolumn{4}{|c|}{NIL} \\ 
 \hline
\end{tabular}
\label{table7}
\centering
\end{table}

With all the data quality analysis performed on both \emph{Questionnaire 1} and \emph{Questionnaire 2} above, 
we have following task design takeaways:
\begin{itemize}
\item Use the dropdown menu whenever possible. In \emph{Questionnaire 1}, we ask several demographic questions. Most
of them are designed to be open-ended questions. For example, one question is to ask about the worker's age. Most of
the workers answer the question properly. However, there is one worker put extra quotation marks around the input number
because we give \emph{(e.g. ``30")} as an example to the question without explicitly stating that quotation marks should not be
typed as part of the answer. The worker is so careful with his response and he follows our instruction word by word. His
dedication leads to an extra effort on the data cleaning.
\item Put quality check on the text responses. In the above analysis, we find that 16\% of text responses from \emph{Questionnaire 1}
do not form 
a complete sentence and some workers copy and paste their answers everywhere. Thus, we may want to put some quality check
on the place where workers need to enter text. One technique is to enforce the minimum length of words for a response and the 
other is to enforce that each response provided by the same worker has to be different.
\item Remove redundant questions. In \emph{Questionnaire 1}, we ask the workers to first pick the bias type 
that they think their responses belong to and then we ask them to write out their responses. However, by carefully examining
those responses, we find that we can easily infer those bias categories. For example, 
``woman do all the housework and raise the kid'' relates to \emph{Gender}. Thus, we think the additional bias category selection
is redundant, which can be removed from the HIT.
\item Detailed instruction provided with examples. In \emph{Questionnaire 2} design, we give workers a concrete setting about finding bias
in ``an unbias image data set'' and instead of one example listed in \emph{Questionnaire 1}, we provide five examples formatted as bullet points. Those instructions can give workers a concrete scenario to 
work with and a few examples give workers ideas of what types of answers in what format we are specifically looking for.
\end{itemize}

\subsection{Demographics}

\begin{figure*}[ht] 
  \begin{subfigure}[b]{0.5\linewidth}
    \centering
    \includegraphics[width=0.75\linewidth]{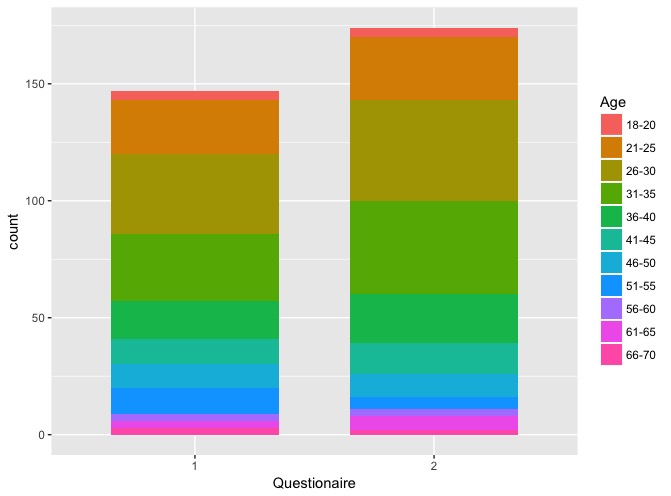} 
	\caption{Age Composition between Two Questionnaires}
    \label{cat-age}
    \vspace{4ex}
  \end{subfigure}
  \begin{subfigure}[b]{0.5\linewidth}
    \centering
    \includegraphics[width=0.75\linewidth]{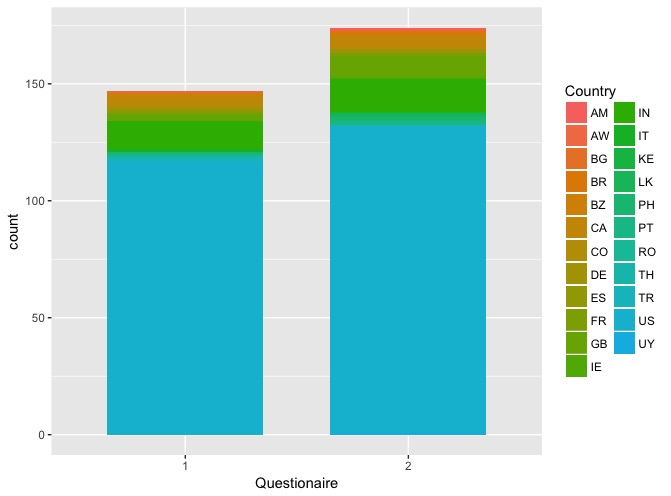} 
	\caption{Country Composition between Two Questionnaires}
    \label{cat-country} 
    \vspace{4ex}
  \end{subfigure} 
  \begin{subfigure}[b]{0.5\linewidth}
    \centering
    \includegraphics[width=0.75\linewidth]{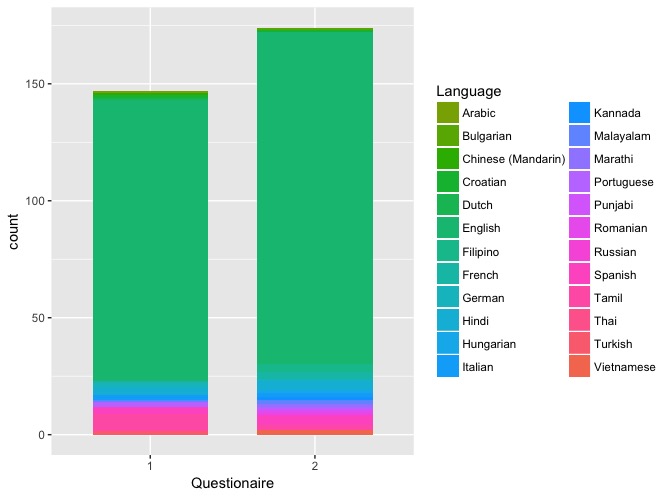} 
\caption{Language Composition between Two Questionnaires}
				\label{cat-lang}
  \end{subfigure}
  \begin{subfigure}[b]{0.5\linewidth}
    \centering
    \includegraphics[width=0.75\linewidth]{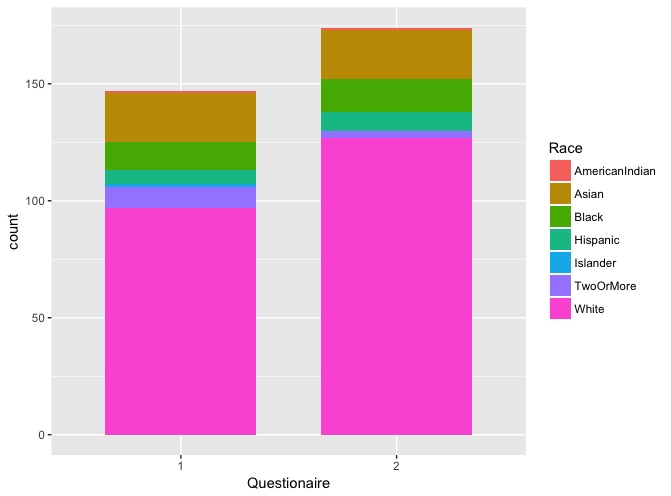} 
\caption{Race Composition between Two Questionnaires}
				\label{cat-race}
  \end{subfigure} 
        \caption{Demographics of the crowd}\label{fig:demographics} 
\end{figure*}

We have a diverse workforce: workers from 17 countries participate in our HIT and they speak. They are originally from 17 different countries and they speak 24 different languages. However, 
certain language, country, and race workers dominate others: 81.60 \% of the population are native English speaker; 75.29 \% of our workforce live in the United States; and 72.99 \% of worker population identify themselves as white. 
The gender ratio is also not balanced: 68.39 \% are female workers. Nonetheless, the age of workers are almost normally distributed, which can be seen in Figure \ref{cat-age}. 
Additional crowd demographic information shown in \ref{cat-country}, \ref{cat-lang}, and \ref{cat-race}. 
Surprisingly, workers from India are fewer than expectation. In addition, the compositions of the workers measured by ``age'', ``country'', ``language'', and ``race'' are similar across
two questionnaires. This finding shows that the crowd demographic is relative stable.

\subsection{Diversity of crowd suggestions }

To study the bias in the crowd, one important question we want to ask is how diverse the crowd's bias is. Specifically, we want
to know: Are a large percentage of workers choosing to suggest stereotypes within a certain category?
How much repetition is there in the suggestions from the crowd? How specific and creative are the suggestions? 
To address those questions, we first count the number of responses under each predefined category. 
As shown in Table \ref{table1}, workers tend to suggest the \emph{Gender} bias, which accounts for 41\% (212) of total responses in \emph{Questionnaire 1}.
Bias in \emph{Race} comes to second with 33\% (170) of total responses. In \emph{Questionnaire 1}, those two categories combined together take 71\% of total 
responses. The dominate share taken by both \emph{Gender} and \emph{Race} is unchanged after we remove all the suspicious responses, which account for 41\% (180) and 32\% (142) of total responses (437) respectively. 

\begin{figure}
\centering
\includegraphics[scale=0.4]{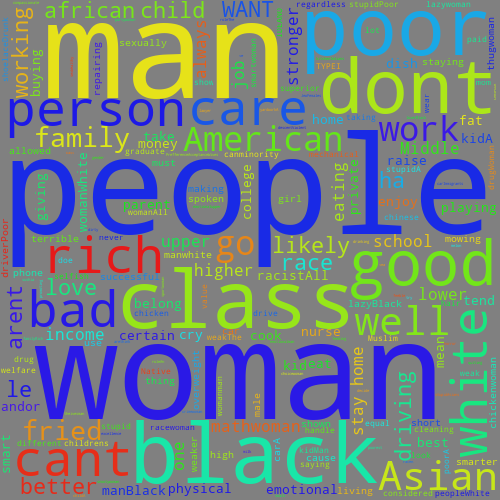} 
\caption{Word cloud of \emph{Questionnaire 1} responses}
\label{fig:wordcloud}
\end{figure}

To better understand the amount of repetition
in the crowd suggestion, we count the frequency of each word and plot the word cloud shown in Figure \ref{fig:wordcloud}. We perform stemming and lemmatization before calculating any statistics related to words. Workers use ``women"
and ``woman'' at the same time when they talk about the gender bias. However, those words should be considered as the same word
when we count the word frequency. We select the high frequency words appeared in the word cloud as the topic words for each category and 
count the frequency of \emph{meaningful} words in the sentences that contain those topic words. 
The result is shown in Table \ref{table4}. \emph{Meaningful} is defined as
the words relate to the \emph{Gender}, \emph{Race}, and \emph{Class} categories. Most of them are noun, adjectives, and verbs. Words like 
``are'', ``all'', ``don't'' are not considered as \emph{meaningful}. In addition, we combine the counts of the words that have similar
meaning together (i.e., ``rich'' and ``wealthy''). Table \ref{table4} sheds some interesting lights on the crowd bias. Workers' bias towards \emph{Gender} is more clustered than towards \emph{Race} and \emph{Class}. For example, the crowd links man with words
``emotion'', ``strong'', ``work'', ``physical'', which are words that often appeared in the bias for man like "Man loves work", 
``Men are physically stronger than women", and ``Men don't like share their emotion". Regarding woman, workers link them with family words
(i.e., ``family", ``children"), certain occupation (i.e., ``nurse''), and certain behaviors (i.e., ``drive''). However, on the other hand,
the crowd has diverse opinions towards the \emph{Race} and \emph{Class}. As can be seen in the Table \ref{table4}, for ``black'',
most of the words have frequency 1. Similar patterns can be seen for ``white'' and ``asian" as well. In addition to the diversity of
the crowd bias towards \emph{Race} and \emph{Class} categories, workers seem to have fewer bias towards ``white'' compared with topic words from other categories. This is probably due to that the majority of the workers answer our HIT task have the race ``white''.
This finding suggests that if researchers want to use the crowd to check any bias in their datasets, they
may want to measure the diversity of the workers bias as well. In our case, lack of bias words associated with topic word ``white``
reflects the possibility of the crowd bias may be dominated by the bias hold by the white people.

To further quantify the diversity of the bias hold by the crowd, we count the number of adjectives that show up in responses less than
three times. This is done with the help of a POS tagger built from NLTK library provided by \citeName{Loper2002}. 
We tag all the words that are adjectives and count the number of words that appeared no more than three times. The heuristic
for this method is that the number of low frequency words for commonly-hold bias for certain topic will be smaller than the 
number of low frequency words for more diverse bias towards a topic. In other words, if people have diverse biases towards a topic,
they may use a variety of words to describe them and for each word, number of times they show up will be low.
The result is shown in Table \ref{table5}. One can see that the adjectives appeared in the sentences topic word ``white''
is zero. This indicates that workers' bias towards "white" are highly clustered. On the other hand, workers may have a much wider diverse bias
towards \emph{Gender} and ``black".

\begin{figure}
\centering
\includegraphics[scale=0.4]{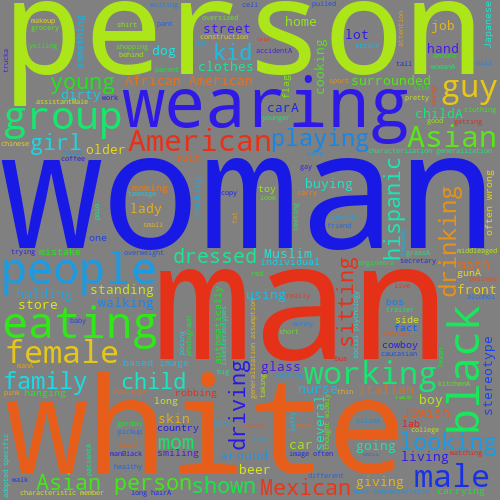} 
\caption{Word cloud of \emph{Questionnaire 2} responses}
\label{fig:wordcloud2}
\end{figure}

We apply the same analysis towards \emph{Questionnaire 2}. The major difference between \emph{Questionnaire 2} and \emph{Questionnaire 1} is that
we remove the category dropdown menu from \emph{Questionnaire 2}. The idea is to give workers more freedom of coming up diverse bias.
The word cloud for \emph{Questionnaire 2} is shown in Figure \ref{fig:wordcloud2}. One can observe that the word frequency related to ``class''
is smaller than the one in \emph{Questionnaire 1}. This observation can be confirmed by the word frequency counts related to each
topic word shown in Table \ref{table8}. Here, workers' responses barely contain word ``poor'' or ``rich'', which are strong indicators
of bias related to ``class''. However, on the other hand, workers focus on the bias related to \emph{Gender} and \emph{Race}. For ``man'', 
the words relate to career (i.e., ``work'', ``boss'') and physical power (i.e., ``construction'' and ``strong''). However, workers
in \emph{Questionnaire 2} link ``man'' with more race-specific words compared with \emph{Questionnaire 1} responses. For example, besides ``black''
and ``white'' shown in \emph{questionnaire 1}, workers also mention the ``hispanic'' and ``asian'' as well. For woman,
family words (i.e., ``children'') and certain occupation (i.e., ``nurse'') still show up in \emph{Questionnaire 2}. 
However, in \emph{Questionnaire 2}, new words such as ``salons'' and ``muslims'' also appeared. For race, workers describe their responses
in much greater detail. For example, for ``black'', they mention ``basketball'' in \emph{Questionnaire 2} 
instead of the generic term ``sports'' in \emph{Questionnaire 1}. In addition, workers in \emph{Questionnaire 2} offer more diverse responses regarding ``white'' than ones in \emph{Questionnaire 1}. Similar to \emph{Questionnaire 1}, \emph{Gender} still has the most diverse response. There are 14 low frequency words for ``man'' and 21
low frequency words for ``woman'' as shown in Table \ref{table5}.

\begin{table*}
\caption{Word frequency appeared in the responses from \emph{Questionnaire 1}}
\label{table4}
\begin{tabularx}{\textwidth}{@{}l*{8}{C}c@{}}
\toprule
people                               & man                   & woman               & black          & white       & asian     & poor & rich  \\    
\midrule
poor (26)                            & woman (25)            & kids, children (10)  & american (6)   & trash (1)   & math (5)  & lazy (2) &  very (3)\\    
black (25)                           & work, jobs (6)        & care (6) & fried (3) & racist (1)  & good (4)  & buying (2) & class (2)\\  
white (23)                           & black (5)             & work (6)             & chicken (2)    & music (1)   & smart (4) & welfare (1) & upper (1)\\   
rich, wealthy (16)                   & white (3)             & drive (6)           & violence (2)   & mansion (1) & drive (4) & uneducated (1) &  tesla (1)\\   
\addlinespace
class (10)                           & strong (5)            & bad (5)             & watermelon (1) & family (1)  & bad (2)   & ticket (1)  & sushi (1)\\   
lazy (9)                             & emotion (4)            & home (4)            & sports (1)     & country (1) & valedictorian (1) & system (1) & smarter(1) \\  
snobby, selfish, greedy (9)   & family (3)         & kitchen (3)         & smart (1)      & collar (1)  & taxi (1)   & succeed (1) & privilege (1)\\    
Chinese (4)                          & cry (2)             & weak (2)            & skilled (1)    &             & stinky (1) & stupid (1)  & eating (1)\\ 
\addlinespace
lower (4)                            & superior (2)          & stupid (2)          & serious (1)    &             & small (1) & steal (1) & dribing (1) \\
upper (3)                            & repairing (2)         & sex (2)             & poor (1)       &             & slanty (1) & schoolwork (1) & well (1)\\
stupid (3)                           & physical (2)          & nurse (2)           & play (1)       &             & piano (1) & rich (1) & arogant (1) \\
men (3)                              & mechanical (2)        & love (2)            & menial (1)     &             & nissan (1) & milks (1) & acting (1)\\
income (3)                           & yardwork (1)          & house(2)            & foolery (1)    &             & great (1) & lottery (1) & lower (1)\\
\bottomrule
\end{tabularx}
\end{table*} 

\begin{table*}
\caption{Word frequency appeared in the responses from \emph{Questionnaire 2}}
\label{table8}
\begin{tabularx}{\textwidth}{@{}l*{8}{C}c@{}}
\toprule
people                              & man                & woman       & black          & white       & asian     & poor & rich    \\    
\midrule
black (19)                          & woman (51)         & asian (15)  & african (23)   & family (4)  & study (8) & food (2)        & wife (1)\\    
woman (12)                          & black (23)         & nurse (13)  & american (22)  & mom (3)     & math (7)  & southerners (1) & mistress (1)\\  
young (7)                           & white (21)         & black (11)  & eat (5)        & rural (3)   & good (5)  & slavery (1)     & guy (1)\\   
asian (5)                           & work (12)          & work (10)   & watermelon (4) & young (3)   & work (3)  & purchase (1)    &  \\   
\addlinespace
muslim (4)                          & boss (7)           & young (9)   & man (4)        & witches (2) & tolerant (3) & mexican (1)  & \\   
indian (4)                          & young (6)          & cook (6)    & chicken (4)    & trash (2)   & shrewd (3)   & junk (1) &  \\  
hair (4)                            & play (5)           & children (6)& basketball (4) & trailer (2) & revealed (3) &  &\\    
terrorist (3)                       & construction (5)   & white (5)   & rap (4)        & smile (2)   & reserved (3) &  & \\ 
\addlinespace
poor (3)                            & asian (5)          & eat (5)     & play (3)       & single (2)  & impatient (3) &  &\\
rich (3)                            & strong (4)         & traditional (3) & pant (3)   & guys (2)    & generous (3) &  & \\
jewish (3)                          & mowing (4)         & salons (3)      & government (3)  & family (2) & friendly (3) &  &  \\
lefties (3)                         & hispanic (4)       & muslim (3)      & thug (2)     & couple (2)    & driver (3) &  & \\
african (3)                         & eat (4)            & italian (3)     & gun (2)    &   american (2)  & domineering (3) &  & \\
\bottomrule
\end{tabularx}
\end{table*}

\begin{table}
\captionsetup{size=footnotesize}
\caption{Number of adjectives with counts no more than three in \emph{Questionnaire 1}}
\label{table5}
\footnotesize\centering

\smallskip 
\begin{tabular*}{\columnwidth}{@{\extracolsep{\fill}}ccc}
\toprule
Topic Word& \emph{Questionnaire 1} Counts & \emph{Questionnaire 2} Counts\\
\midrule
man & 18 & 14   \\
woman & 10 & 21\\
black & 7 & 5 \\
white & 0 & 5 \\
asian & 2 & 3 \\
poor & 3 & 1\\
rich & 1 & 0\\
\bottomrule
\end{tabular*}
\end{table}

\subsection{Deeper Dive into Questionnaire 1}
The results from \emph{Questionnaire 1} made it clear that there were problems with the HIT design. The respondents almost always gave vague phrases, or sometimes only a word, instead of scenarios that could be captured in a photograph. This revealed a key ambiguity in our task design. Although the examples given in the instructions were scenarios, the instructions never explicitly said that a visual scenario was the desired suggestion type. This was one key change we realized was necessary for \emph{Questionnaire 2}. However, the ambiguity of the instructions did offer one potential benefit in an extra area of analysis. Who were the respondents that were able to intuit what we wanted, even with poor instructions? And what do they have in common? 

With these questions in mind, the 147 crowd responses from \emph{Questionnaire 1} were narrowed down to only include responses that contained one or more suggestion that was either phrased in the way that presented a scenario, or included stereotypes that were more creative or less obvious. This narrowed the data set down to 44 responses. Some of these can be seen in Table 7. Choosing the ``best" suggestions was a subjective task, and due to the repetition it was often random which of a repeated suggestion was chosen for the reduced data set.  

Contrary to what we had hoped to find, the demographics of these 44 respondents were found to be fairly representative of the total group. 81\% spoke English as a native language, 63\% were currently living in the US, and 68\% were women. While the demographics of this small group did not ultimately reveal much about an ideal workforce, observing the best suggestions did point out some other interesting observations.

Within these 44 best suggestions, it was often the case that two of the suggestions would be poor with only one being of higher quality. For example, one respondent suggested the scene of woman feeding a baby. It is a common stereotype that women do the extra labor in taking care of children, so this was counted as a good suggestion despite it being fairly common. However, another response by this respondent was ``A Bird sheltering her babies.'' What the respondent meant by this is unclear, and it highlights the extreme variability in quality of answers, even from the same worker. 

A few of the suggestions also revealed that cultural differences did broaden the suggestions we received. For example, one respondents made suggestions such as ``Women's are not allowed to go outside after 8:00 PM" and ``Women's should not go to Parties," which are stereotypes that have diminished in most of western culture but should not be ignored in a data set that is representing many societies. 

\begin{table}
\centering
\caption{Some of the suggestions from the crowd responding to Questionnaire 1}
\begin{tabular}{ |p{2cm}||p{5cm}|  }
 \hline
 \multicolumn{2}{|c|}{\textbf{Crowd Suggestions} } \\
 \hline
 \textbf{Category} & \textbf{Suggestion} \\
 \hline
 Class   & The poor person is buying processed food   \\
 \hline
 Gender & The pilot on the airplane is a man. \\
 \hline
 Race & A black person being shown as a cleaner/janitor/other menial role \\
 \hline
 Race & An Asian family being shown driving a Nissan. \\
 \hline
 Class & A poor person buying lottery tickets. \\
 \hline
 Gender & Men enjoy metal music. \\
 \hline
 Other & Teens texting on their phone walking across roads \\

 \hline
\end{tabular}
\centering
\end{table}

\subsection{Deeper Dive into Questionnaire 2}

Altering the structure of \emph{Questionnaire 2} resulted in more stereotypes in the form of scenarios that could be photographed. However, while the quality improved somewhat, there was still a large amount of repetition in the workers' suggestions. There were repeated cases where workers suggested stereotypes that were explicitly listed in the instructions as stereotypes we were already aware of and did not want repeated. Even when workers followed the instructions, and clearly tried to be somewhat creative, there still seemed to be a disconnect about what would constitute a helpful suggestion. 

In an attempt to quantify whether removing the categories from the second HIT increased the diversity of the responses, we manually went through all the stereotypes from the crowd and labeled them based on what category they would fall under. As we've progressed in this research, it has become clear that many stereotypes can not be limited to just one category. For example, one worker wrote ``A white man in an astronaut suite.'' This stereotype reflects both the history of women and people of color having limited access to important roles in the sciences, and therefore falls under both the gender and race categories. Therefore, the category counts from \emph{Questionnaire 2} were not as clear cut as \emph{Questionnaire 1}. However, of the 407 crowd suggestions that provided enough information to classify them, 157 were about race, 105 were about gender, 25 were about class, and 120 were about various other scenarios. These numbers show a significant percentage increase in the ``other" category. However, a noticeable amount of the stereotypes in the ``other" category were useless with examples like: ``Waiting in line for coffee.'' The race and gender stereotypes given, while often repetitive and uncreative, were typically fairly concrete. Some examples of helpful increases within the ``other" category were those surrounding sexuality and disability. 

In order to try and get a better grasp on what category combinations and new categories were represented, we ran a sentence clustering algorithm from RxNLP's NLP library\footnote{http://www.rxnlp.com/}. The 450 stereotypes from workers were passed to this algorithm and clustered into 37 categories. Some of the resulting clusters were broad, such as those containing all cases of stereotypes surrounding Americans, including African Americans. However, the clusters did point out a number interesting commonalities, including the surprising number of stereotypes surrounding hair: ``A person who's wearing dreads smoking weed'', ``Man shown doing a really bad job of braiding his daughter's hair,'' and numerous examples surrounding black women's hair. 

There were other similarly clustered responses around specific aspects of a person's appearance, such as glasses. The idea that glasses imply someone is a nerd, or studious, was very common. Comments about millennials being obsessed with technology were also prevalent. 

Within the workers who seemed to understand the task, there was a wide range of answers. Some workers clearly reflected on their personal experience. For example, one worker gave multiple suggestions about left handed individuals such as ``Left-handed individual being scolded for using machinery incorrectly." Meanwhile, other workers gave stereotypes that made us reconsider what we were hoping to find. For example, one worker gave multiple stereotypes surrounding dogs such as ``A husky shown in a film as a wolf" with the explanation that ``Most people believe that husky's are of wolf descent, but the truth is they contain very little to no wolf in them." At first thought, this seemed useless. However, this worker was thinking outside of the box which is ultimately what we asked for. Perhaps this input could be used for a dataset that was concerned with identifying dog species. Table 8 displays some of the other suggestions from \emph{Questionnaire 2}. 

Finally, one worker provided some creative suggestions, but framed their responses as images that would counteract the stereotype rather than reinforce it. Their suggestions can be seen in Table 9, and made us consider that in future work we could ask for image suggestions that subvert stereotypes rather than those that reinforce them. We believe this would naturally reduce some of the repetition in worker responses, because even if workers thought of the same stereotype, their suggestions of ways to subvert it would likely be different. 

\begin{table}
\centering
\caption{Some of the suggestions from the crowd responding to Questionnaire 2}
\begin{tabular}{ |m{6cm}|  }
 \hline
 \hline
 \textbf{Crowd Suggestions} \\
 \hline
  Manly looking transgender woman  \\
  \hline
  A white tv jourlanist \\
  \hline
  Micronesian family with many children and a small house. \\
  \hline
  A "nerdy" white male in his twenties codin \\
  \hline
  A white man in an astronaut suite.\\
  \hline
  Male pastor\\
  \hline
  A white rural American sitting on a front porch with a Confederate flag hanging nearby.\\
 \hline
\end{tabular}
\centering
\end{table}

\begin{table}
\centering
\caption{One worker framed their suggestions as if they were proposing images that would counteract stereotypes and led us to consider that strategy for future work. }
\begin{tabular}{ |m{6cm}|  }
 \hline
 \hline
 \textbf{One worker's responses} \\
 \hline
 A heavier weight male or female hiker, backpacker, or mountain climber. \\
  \hline
  A woman working in the maritime industry (driving a vessel, hoisting the sails etc) \\
  \hline
  Millenial in a position of authority/management over a baby boomer or gen x'er. \\
  \hline
  \end{tabular}
\centering
\end{table}

\section{Discussion}
\label{section:discussion}

The key question we ask ourselves is: “Did the outcome of this research support the idea that harnessing the diversity of the crowd could be useful for anticipating bias in data sets?” Our answer is that the crowd demonstrated some modest abilities in being useful at this task, but this method would require a research team that is highly motivated to weed out bias in a data set and willing to invest in a fairly time consuming and tedious task. The two HITs yielded some valuable insights from the crowd, but it was far more common for a worker to provide a useless or very common stereotype. This made quality control difficult. Determining how to accept or reject worker suggestions is problematic as it is both time consuming, and sometimes difficult to ascertain whether the worker is willfully avoiding completely the task correctly or whether they are just unequipped for the task. Our strategy was to simply accept all responses and then wade through them to search for the rare valuable insights. 

The results of our two HITs suggest that the crowd struggled with the level of freedom and ambiguity inherent to our task. Some of the poor results were due to bad worker behavior, i.e. ignoring some of our explicit instructions, but other workers who gave poor responses seemed to be trying based on their justifications in \emph{Questionnaire 2}. It would have been easier to ask more straightforward questions about stereotypes, like the work of \citeauthor{bolukbasi2016man} \shortcite{bolukbasi2016man}, however we explicitly wanted to test whether the crowd could be used in a more predictive way rather than evaluating potential problems we are already aware of. 

At the outset of this project, we feared receiving openly malicious or ugly comments from the crowd, but thankfully found that not to be the case. There were no truly malicious comments, although some workers clearly seemed to care about our motivations more than others. 

The problem of bias in machine learning is a complex issue and we hope that this research will add to the discussion. While we would not yet recommend our exact strategy to other researchers, we hope that this experiment could help point out some of the inherent difficulties of using the crowd in this challenging, but important space. 

\section{Future Work}
\label{section:future}

There have been many challenges for crowdsourcing work. 
Specifically for our task, one challenge comes from the data quality check and cleaning.
In this paper, we implement a few simple techniques on data quality assurance. 
However, those techniques may still suffer from lacking of accuracy
especially when we use them in a large scale. One issue is that we do not value the response from the semantics perspective. Workers may
enter the responses that are unique and more than three words but the semantics of those responses are irrelevant to the task goal
(one example is ``I WANT A CLASS TYPE"). We find that
manually checking the responses are still needed for quality assurance purpose. Clearly, manual examination cannot scale well when we collect
thousands of responses. More advanced NLP techniques such as semantic parsing might be needed in order to build a more accurate
quality assurance tool for text data that can perform relative well in a large scale. In addition, we may want to implement the quality
check for each text response.




\bibliographystyle{aaai}
\bibliography{bib}

\end{document}